\documentclass[twocolumn,floatfix, nofootinbib,prd,preprintnumbers,showpacs,showkeys,superscriptaddress,longbibliography]{revtex4-1}

\usepackage[mathscr]{euscript}
\usepackage{amsmath}
\usepackage{graphicx}
\usepackage{dcolumn}
\usepackage{bm}
\usepackage{epsfig}
\usepackage{amssymb,latexsym,mathrsfs}
\usepackage{graphicx}
\usepackage{color}
\usepackage{hyperref}
\usepackage{float}
\usepackage[thinlines]{easytable}
\usepackage{multirow}
\usepackage{diagbox}

\newcolumntype{M}[1]{>{\centering\arraybackslash}m{#1}}
\newcolumntype{N}{@{}m{0pt}@{}}

\usepackage{tikz}
\usepackage[compat=1.1.0]{tikz-feynman}

\usepackage{amsmath}
\newcommand*\diff{\mathop{}\!\mathrm{d}}

\hypersetup{
    colorlinks=true,
    linkcolor=red,
    citecolor=blue,
}

\usepackage{amsmath}
\usepackage{amssymb}
\usepackage{subfigure}
\usepackage{hyperref}
\usepackage{url}
\usepackage{xcolor}
\usepackage{color}
\definecolor{amaranth}{rgb}{0.9, 0.17, 0.31}
\definecolor{purple(munsell)}{rgb}{0.62, 0.0, 0.77}
\definecolor{americanrose}{rgb}{1.0, 0.01, 0.24}
\definecolor{palatinateblue}{rgb}{0.15, 0.23, 0.89}
\definecolor{royalblue(web)}{rgb}{0.25, 0.41, 0.88}
\definecolor{hanpurple}{rgb}{0.32, 0.09, 0.98}
\definecolor{beaublue}{rgb}{0.74, 0.83, 0.9}
\definecolor{carminered}{rgb}{1.0, 0.0, 0.22}
\definecolor{brightpink}{rgb}{1.0, 0.0, 0.5}
\definecolor{vividviolet}{rgb}{0.62, 0.0, 1.0}

\definecolor{electron}{rgb}{1.0, 0.67, 0.22}
\hypersetup{ linktoc=all,
    colorlinks, linkcolor={palatinateblue},
    citecolor={brightpink}, urlcolor={amaranth}}

\newcommand{\be}{\begin{equation}}
\newcommand{\ee}{\end{equation}}
\newcommand{\bs}{\begin{split}} 
\newcommand{\bea}{\begin{eqnarray}}
\newcommand{\eea}{\end{eqnarray}}

\newcommand{\bes}{\begin{subequations}}
\newcommand{\ees}{\end{subequations}}

\newcommand{\bo}{\raise-1mm\hbox{\Large$\Box$}}

\begin{document}
\title{Electron as a Tiny Mirror: Radiation From a Worldline With Asymptotic Inertia}

\author{Michael R.R. Good}
\email{michael.good@nu.edu.kz}
\affiliation{Department of Physics \& Energetic Cosmos Laboratory, Nazarbaev University,\\ Astana 010000, Qazaqstan}
\affiliation{Leung Center for Cosmology $\&$ Particle Astrophysics,
National Taiwan University, \\ Taipei 10617, Taiwan}
	
\author{Yen Chin \surname{Ong}}
\email{ycong@yzu.edu.cn}
\affiliation{Center for Gravitation and Cosmology, College of Physical Science and Technology, Yangzhou University, \\Yangzhou, 225002, China}
\affiliation{Shanghai Frontier Science Center for Gravitational Wave Detection, School of Aeronautics and Astronautics, Shanghai Jiao Tong University, \\ Shanghai 200240, China}

\begin{abstract}
We present a moving mirror analog of the electron, whose worldline possesses asymptotic constant velocity with corresponding beta Bogolubov coefficients that are consistent with finite total emitted energy. Furthermore, the quantum analog model is in agreement with the total energy obtained by integrating the classical Larmor power. 
\end{abstract} 

\maketitle
\section{Introduction: Fixed Radiation}
Uniform acceleration, while beautiful and simple is not globally physical.  Consider the problem of infinite radiation energy from an eternal uniformly accelerated charge.  The physics of eternal unlimited motions is not only the cause of misunderstandings, but also the starting point of incorrect physical interpretations, especially when considering global calculable quantities, like the total radiation emitted of a moving charge. Infinite radiation energy afflicts accurate scrutiny of physical connections between acceleration, temperature, and particle creation. 

More collective consideration should be prescribed to straighten out the issue.  One path forward, is the use of limited non-uniform accelerated trajectories, capable of rendering finite global total radiation energy. The trade-off with these trajectories is usually the lack of simplicity or tractability in determining the radiation spectrum in the first place.    

In this short Letter, we present a solution for finite radiation energy and its corresponding spectrum. Limited solutions of this type are rare and can be employed to  investigate the physics associated with contexts where a globally continuous equation of motion is desired. For instance, the solution is suited for applications like the harvesting of entropy from a globally defined trajectory of an Unruh-DeWitt detector, the non-equilibrium thermodynamics of the non-uniform Davies-Fulling-Unruh effect, or the dynamical Casimir effect \cite{moore1970quantum}, and particle production of the moving mirror model \cite{DeWitt:1975ys,Davies:1976hi,Davies:1977yv}. 

Providing a simple conceptual and quantitative analog application to understanding the radiation emitted by an electron, we demonstrate the existence of a correspondence (see similar correspondences in \cite{Nikishov:1995qs,Ritus:2003wu,Ritus:2002rq,Ritus:1999eu,Good:2022gvk,Ritus:2022bph,Good:2022eub}) between it and the moving mirror.  At the very least, this functional coincidence is general enough to be applied to any tractably integrable rectilinear classical trajectory that emits finite radiation energy.  Here, we analytically compute the relevant integrable quantities for the specific solution and demonstrate full consistency.  The analog approach treats the electron as a tiny moving mirror, somewhat similar to the Schwarzschild \cite{Good:2016oey}, Reissner–Nordstr\"om \cite{good2020particle}, and Kerr \cite{Good:2020fjz} black mirror analogies, but with asymptotic inertia of a limited acceleration trajectory. Interestingly, the analog reveals previously unknown electron acceleration radiation spectra, thus helping to develop general but precise links between acceleration, gravity, and thermodynamics.  

\section{Elements of Electrodynamics: Energy From Moving Electrons}
In electrodynamics \cite{Jackson:490457, Zangwill:1507229, Griffiths:1492149}, the relativistically covariant Larmor formula (the speed of light $c$, the electron charge $q_e$, and vacuum permittivity $\epsilon_0$ are set to unity),
\be P = \frac{\alpha^2}{6\pi},\ee
is used to calculate the total power radiated by a point charge as it accelerates \cite{Jackson:490457}. Its usefulness is due in part to Lorentz invariance and the fact that proper acceleration, $\alpha$, is intuitive, being what an accelerometer measures in its own instantaneous rest frame \cite{Rindler:108404}. 

When any charged particle accelerates, energy is radiated in the form of electromagnetic waves, and the total energy of these waves is found by integrating over coordinate time.  That is, the integral
\be \label{Elarmor} E = \int_{-\infty}^{\infty} P \diff{t},\ee
demonstrates that the Larmor power, $P=\alpha^2/6\pi$, directly tells an observer the total energy emitted by a point charge along its time-like worldline. This includes trajectories that lack horizons, see e.g., \cite{Good:2016atu}.  This result is finite, only when the proper acceleration is asymptotically zero; i.e. the worldline must be asymptotically inertial.

The force of radiation resistance, whose magnitude is given relativistically as the proper time derivative of the proper acceleration,
\be F = \frac{\alpha'(\tau)}{6\pi}, \label{force}\ee
is known as the magnitude of the Lorentz-Abraham-Dirac (LAD) force, e.g. \cite{Myrzakul:2021bgj}.  The power, $F \cdot v$, associated with this force can be called the Feynman power \cite{Feynman:1996kb}.  The total energy emitted is also consistent with the Feynman power, where one checks:
\be E = -\int_{-\infty}^{\infty} F\cdot v \diff{t}.\label{Efeynman}\ee
The negative sign demonstrates that the total work against the LAD force represents the total energy loss.  That is, the total energy loss from radiation resistance due to Feynman power must equal the total energy radiated by Larmor power.  Larmor and Feynman powers are not the same, but the magnitude of the total energy from both are identical, at least for rectilinear trajectories that are asymptotically inertial.   

Interestingly, the above results also hold in a quantum analog model of a moving mirror. 
A central novelty of this work is to explicitly connect the quantum moving mirror radiation spectra with classical moving point charge radiation spectra.  
Traditionally, e.g. \cite{DeWitt:1975ys,Davies:1976hi,Davies:1977yv,walker1985particle,carlitz1987reflections,Ford:1982ct}, and recently, e.g. \cite{Good:2019tnf,Moreno-Ruiz:2021qrf,Good:2020fsw}, moving mirror models in $(1+1)$-dimensions are employed to study properties of Hawking radiation for black holes. Here, we show that it is also useful to model the spectral finite energy of electron radiation. In particular, a suitably constructed mirror trajectory (which is quite natural) can produce the same total energy consistent with the above via
\be \label{Ebeta} E = \int_0^\infty \int_0^\infty \omega |\beta_{\omega\omega'}|^2 \diff{\omega}\diff{\omega'}.\ee
The final drifting speed of the mirror or electron will be less than the speed of  light, labeled $s$, with $0<s<1$. We also denote $a:= \omega(1+s) + \omega'(1-s)$, $b:=\omega(1-s)+\omega'(1+s)$, and $c:=a+b$; $d:=a-b$. Note that $c = 2(\omega + \omega')$.   

\section{GO Trajectory for Finite Energy Emission}
\begin{figure}[htbp]
\centering
  \centering
  \includegraphics[width=0.9\linewidth]{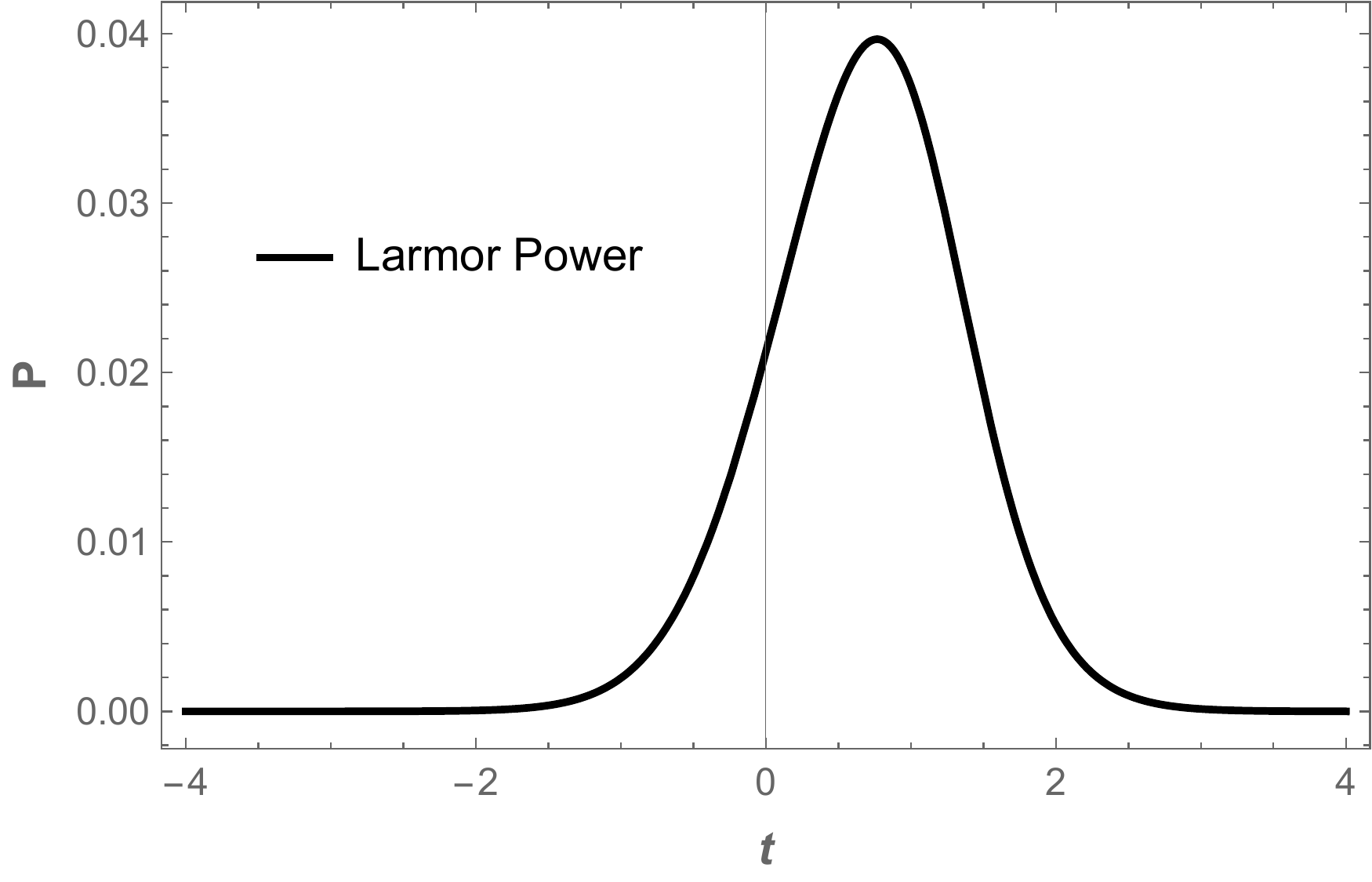}
 \caption{A plot of the Larmor power, Eq.~(\ref{power}), of the GO trajectory, Eq.~(\ref{traj}), as a function of time and final constant speed, $s=0.9$, i.e. Eq.~(\ref{powerGO}). Here $\kappa = 1$. This plot helps illustrate that the Larmor power never emits negative energy flux (NEF) and asymptotically dies off, consistent with a physically finite amount of total radiation energy, Eq.~(\ref{Elarmor}).  }
\label{power}
\end{figure}

We consider a globally defined, continuous worldline, which is rectilinear, time-like, and possesses asymptotic zero velocity in the far past, while travelling to asymptotically constant velocity in the far future (but asymptotically inertial both to the past and the future).  
It radiates a finite amount of positive energy and has beta Bogolubov coefficients that are analytically tractable.   
The `GO' trajectory, if you will,  goes like (Good-Ong 2015 \cite{Good:2015nja}), 
\begin{equation}\label{traj}
z(t) = \frac{s}{2 \kappa} \ln(e^{2 \kappa t}+1),
\end{equation}
where $\kappa$ is an acceleration parameter.
Its total power, if we applied the Larmor formula $P =\alpha^2/6\pi$, is
\be P = \frac{ 2 \kappa ^2}{3 \pi}\frac{ s^2 e^{-4 \kappa  t}\left(1+e^{-2 \kappa  t}\right)^2}{\left[\left(1+ e^{- 2 \kappa t}\right)^2-s^2\right]^3}.\label{powerGO}\ee
Notice the power is always positive and asymptotically drops to zero.  See Figure \ref{power} for an illustration. 

\begin{figure}[htbp]
\centering
  \centering
  \includegraphics[width=0.9\linewidth]{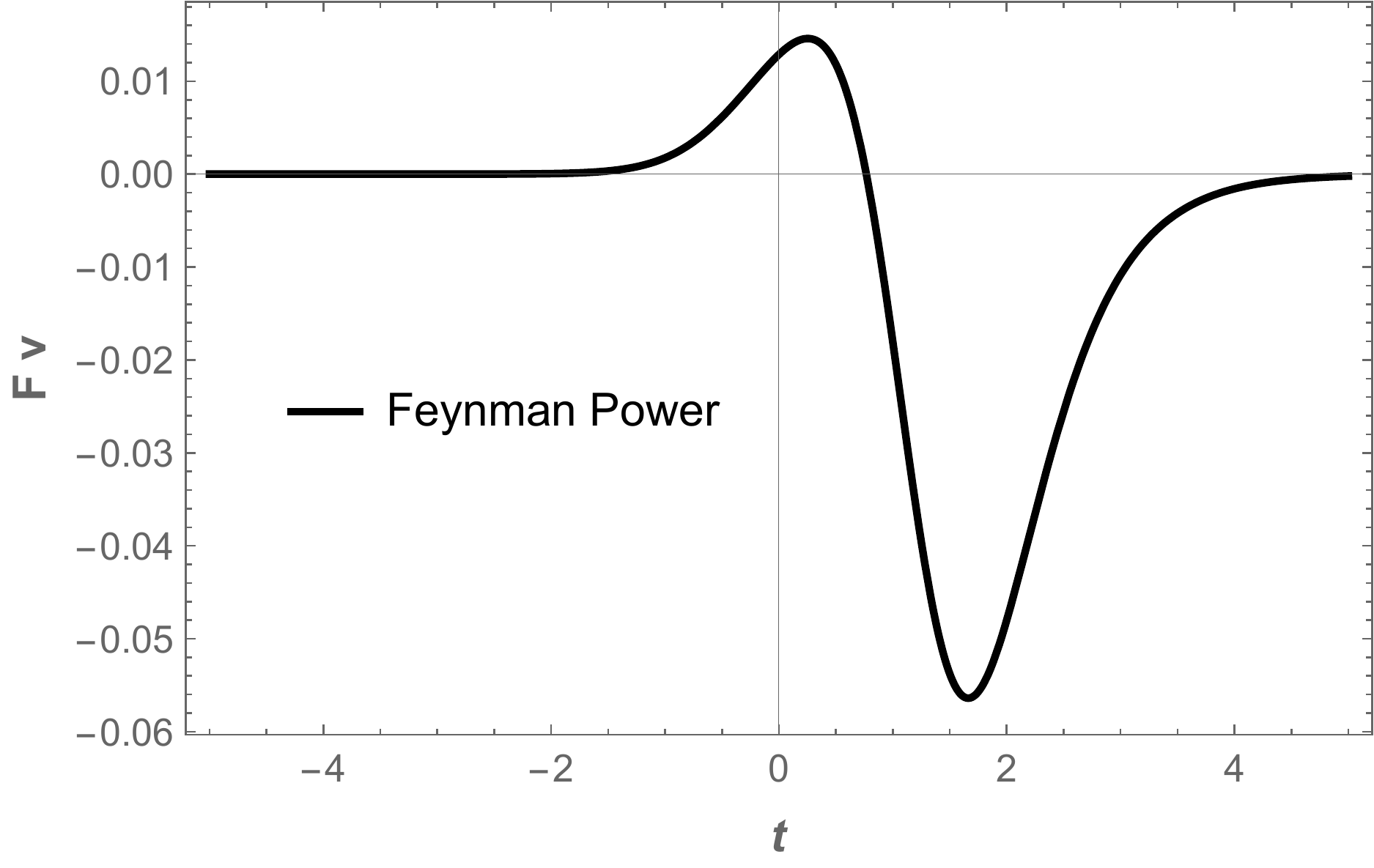}
 \caption{A plot of the Feynman power, $F\cdot v$ associated with the self-force Eq.~(\ref{force}), of the GO trajectory, Eq.~(\ref{traj}), as a function of time and final constant speed, $s=0.9$, i.e. Eq.~(\ref{FpowerGO}). Here $\kappa = 1$. This plot helps illustrate the Feynman power dies off asymptotically, has a period of negative radiation reaction, and is also consistent with a physically finite amount of total radiation energy, Eq.~(\ref{Efeynman}).  }
\label{Fpower}
\end{figure}

The Feynman power, $F\cdot v$, associated with the self-force Eq.~(\ref{force}) is
\be F\cdot v = \frac{2 \kappa ^2 s^2 e^{4 \kappa  t} \left(j_1 e^{6 \kappa  t}+j_2 e^{4 \kappa  t}+e^{2 \kappa  t}+1\right)}{3 \pi  \left( -j_1 e^{4 \kappa  t}+2 e^{2 \kappa  t}+1\right)^3}\label{FpowerGO}\ee
where $j_1=s^2-1$ and $j_2 = 2s^2-1$. Just like the Larmor power, Eq.~(\ref{powerGO}), the Feynman power, Eq.~(\ref{FpowerGO}), asymptotically dies off, but unlike the Larmor power, the Feynman power has a period of negative radiation reaction.  See Figure \ref{Fpower} for an illustration. 

\begin{figure}[htbp]
\centering
  \centering
  \includegraphics[width=0.9\linewidth]{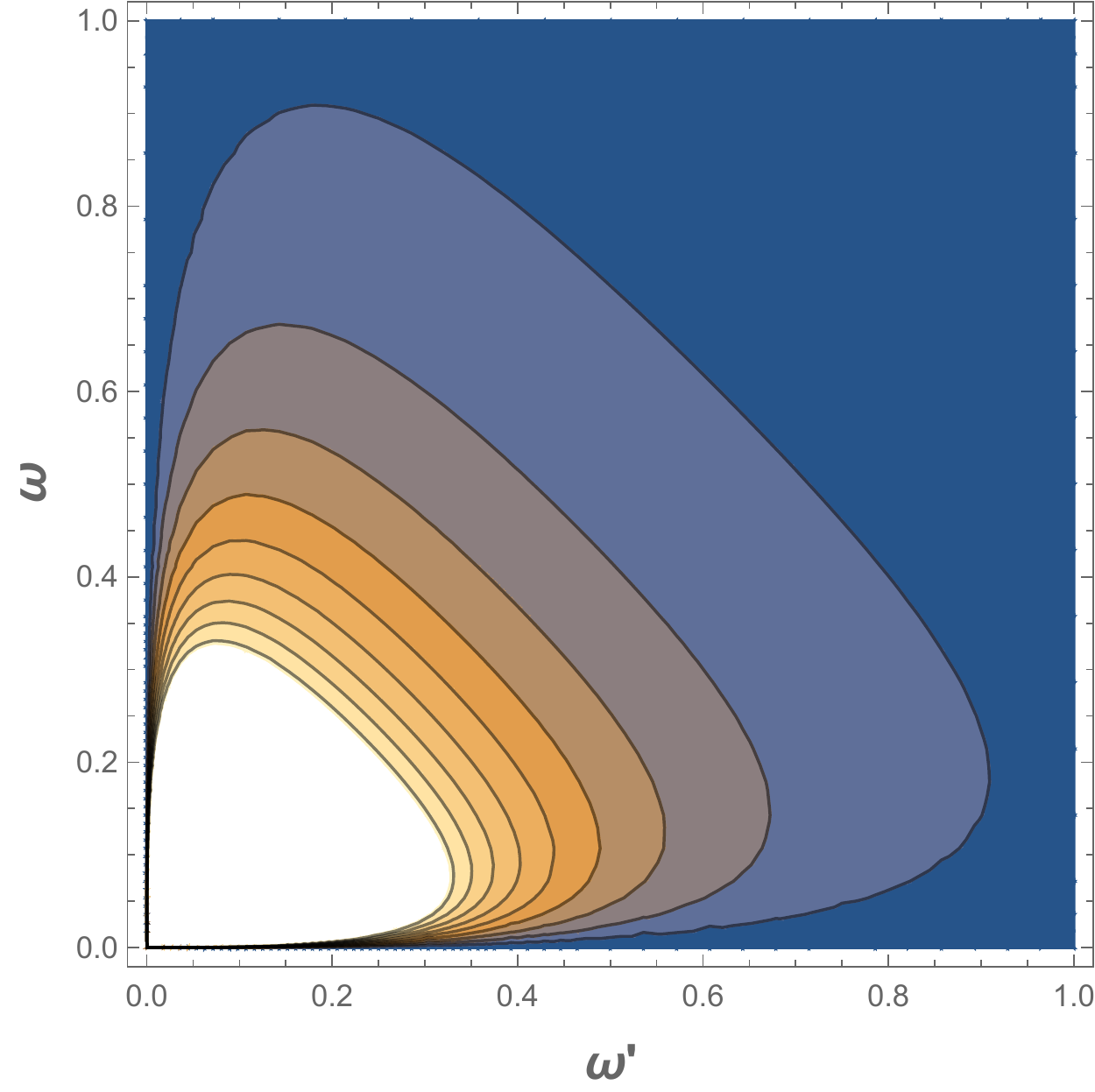}
 \caption{A plot of the coefficients Eq.~(\ref{betas}), as a function of in and out modes, $\omega$ and $\omega'$, where final constant speed, $s = 0.444$ for illustration. Here $\kappa = 1$. This plot underscores the symmetry of the modes in the particle per mode squared distribution spectrum of the beta Bogoliubov coefficients, Eq.~(\ref{betas}).  }
\label{GO_FigC}
\end{figure}

\begin{figure}[htbp]
\centering
  \centering
  \includegraphics[width=0.9\linewidth]{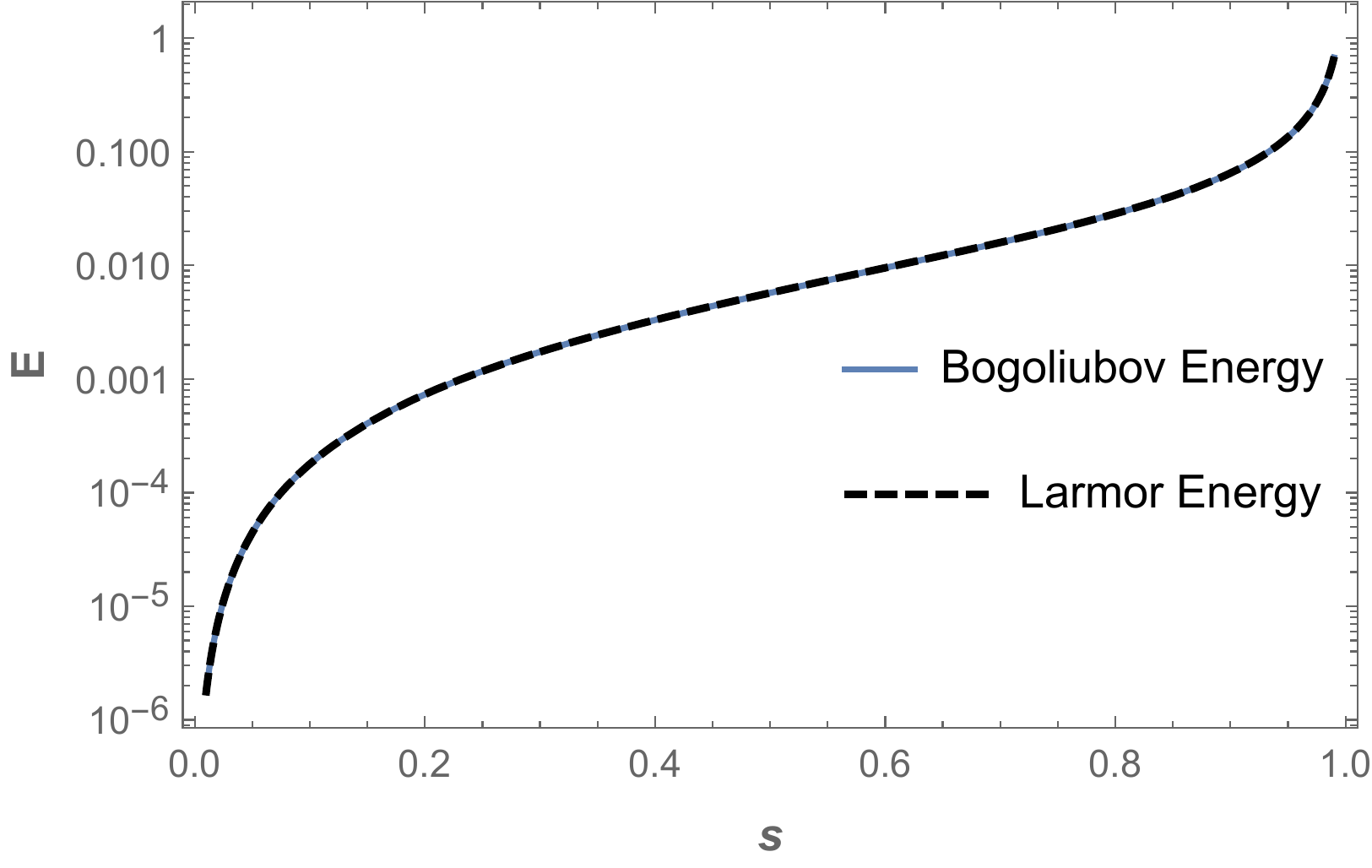}
 \caption{A plot of the Larmor energy, Eq.~(\ref{LarmorEnergy}) and Bogoliubov energy Eq.~(\ref{Ebeta}) using coefficients Eq.~(\ref{betas}), both as a function of final constant speed, $s$.  Here $0<s<0.99$ and $\kappa = 1$. This plot helps confirm that Larmor energy and Bogoliubov energy are equivalent, substantiating the double-sided moving mirror as an analog model of the electron.  }
\label{GO_Fig1}
\end{figure}

We now compute the total energy, using either the Larmor power or Feynman power, and integrating over time.  In terms of the rapidity $\eta = \tanh^{-1}v$ and Lorentz factor $\gamma$, the total energy is given by
\be 
E = \frac{\kappa}{24\pi} \left[ \left(\gamma^2 -1\right) + \left(\frac{\eta }{s}-1\right)\right]. \label{LarmorEnergy}
\ee
We remark that $(\frac{\eta}{s}-1) = \frac{1}{2s}\ln\frac{1+s}{1-s} -1$ is proportional to the lowest order soft energy of inner bremsstrahlung in the case of beta decay (equation 3 in \cite{Good:2022eub}), which is the deep IR contribution.  One can see that Eq.~(\ref{LarmorEnergy}) is finite for all $0<s<1$ and consistent with both the Larmor power and Feynman power.  After we compute the beta Bogoliubov spectrum and plot it in Figure \ref{GO_FigC}, we will compute its total energy and plot it in Figure \ref{GO_Fig1}. We call Eq.~(\ref{LarmorEnergy}), the Larmor energy to differentiate it from the Bogoliubov energy, Eq.~(\ref{Ebeta}) found after substituting Eq.~(\ref{betas}).  The energy is a function of the final constant speed, $s$.

Finally, the spectrum given by the Bogolubov coefficients is best found by first considering the presence of a mirror in vacuum, e.g. \cite{Birrell:1982ix,good2020extreme}.  The mode functions that correspond to the in-vacuum state, 
\be \phi^{\textrm{in}}_{\omega'} = \frac{1}{\sqrt{4\pi \omega'}}\left[e^{-i\omega' v} - e^{-i \omega'p(u)}\right],\ee
and mode functions that correspond to the out-vacuum state,
\be \phi^{\textrm{out}}_{\omega} = \frac{1}{\sqrt{4\pi \omega}}\left[e^{-i\omega f(v)} - e^{-i \omega u}\right],\ee
comprise the two sets of incoming and outgoing modes needed for the Bogolubov coefficients. The $f(v)$ and $p(u)$ functions express the trajectory of the mirror, Eq.~(\ref{traj}), but in null coordinates, $u=t-z$ and $v=t+z$. In spacetime coordinates congruent with Eq.~(\ref{traj}),  one form of the beta integral \cite{Good:2016atu} for one side of the mirror is,
\be \beta_{\omega\omega'} = \int_{-\infty}^{\infty} \diff z \frac{e^{i \omega_n z - i \omega_p t(z)}}{4\pi\sqrt{\omega\omega'}} \left[\omega_p - \omega_n t'(z)\right],\ee
where $\omega_p = \omega + \omega'$ and $\omega_n = \omega-\omega'$. Combining the results for each side of the mirror \cite{Good:2016yht} by adding the squares of the beta Bogolubov coefficients ensures that we account for all the radiation emitted by the mirror \cite{Zhakenuly:2021pfm}. The overall count per mode per mode is
\be |\beta_{\omega\omega'}|^2 = \frac{s^2 \omega  \omega ' Z}{2 \pi  a b c d \kappa}\left(\frac{e^{\frac{\pi  d}{4 \kappa }}-1 }{e^{\frac{\pi  c}{4 \kappa }}-1} \right) e^{\frac{\pi  b}{4 \kappa }},\label{betas}\ee
where $ Z = b \, \text{csch}\left(\frac{\pi  a}{4 \kappa }\right)+a\, \text{csch}\left(\frac{\pi  b}{4 \kappa }\right)$. Eq.~(\ref{betas}) combines the squares, $ |\beta_R|^2+|\beta_L|^2$, of the coefficients for each side of mirror \cite{Good:2015nja}.  See a plot of the symmetry between the modes $\omega$ and $\omega'$ in Figure \ref{GO_FigC}. 

It is then straightforward to verify that the total energy obtained by integrating the power is the same as by using the beta Bogoliubov integral Eq.~(\ref{Ebeta}).  We cannot prove this analytically, but a numerical integral is sufficiently convincing.  See Figure \ref{GO_Fig1} for a plot of the Larmor and Bogoliubov energies as a function of final constant speed.

\section{Discussions: Mirrors, Electrons, and Black Holes}
Prior studies of accelerated electrons and their relationship to mirrors are few, however several works, e.g. \cite{Myhrvold:1983hv,Paithankar:2020akh}, connect electrons to the general Davies-Fulling-Unruh effect (e.g. perhaps the most well-known is Bell and Leinaas \cite{Bell:1982qr} which considered the possibility of using accelerated electrons as thermometers to demonstrate the QFT relationship between acceleration and temperature).  Nevertheless, perhaps an early clue a functional identity existed was made in 1982 by Ford and Vilenkin \cite{Ford:1982ct} who found the LAD self-force was the same form for both mirrors and electrons. In 1995, Nikishov and Ritus \cite{Nikishov:1995qs} asserted the spectral symmetry and found that the LAD radiation reaction has a term that corresponds to the negative energy flux (NEF) from moving mirrors. Ritus examined \cite{Ritus:2003wu,Ritus:2002rq,Ritus:1999eu} the correspondence connecting the radiation from both the electron and mirror systems, claiming not only a deep symmetry between the two, but an fundamental identity related to bare charge quantization \cite{Ritus:2022bph}. Recently, the duality was extended to Larmor power \cite{Zhakenuly:2021pfm} and the deep infrared \cite{Good:2022eub}.  The approach has pedagogical application; for instance, it was used to demonstrate the physical difference between radiation power loss and kinetic power loss \cite{Good:2022gvk}.


The GO moving mirror was initially constructed to model the evaporation of black holes that exhibit a ``death gasp'' \cite{Bianchi:2014vea,Bianchi:2014qua,Abdolrahimi:2015tha} -- an emission of NEF due to unitarity being preserved.
Therefore it is a sturdy result that the total finite energy emitted from the double-sided mirror matches the result from the Larmor formula for an electron. In this sense the GO mirror trajectory is a crude but functional analog of a drifting electron that starts at zero velocity and speeds away to some constant velocity. A single-sided moving mirror does not account for all the radiation emitted, and differs from the electron spectra.  Notably, there is no known NEF radiated from an electron. 

In the literature, one finds some properties that are shared by black holes and electrons. For example, the ratio of the magnetic moment of an electron to its spin angular momentum is $ge/2m$ with $g=2$, which is twice the value of the gyromagnetic ratio for a classical rotating charged bodies ($g=1$). Curiously, as Carter has shown \cite{Carter:1968rr}, a Kerr-Newmann black hole also has $g=2$. This has led to some speculations on whether the electron is a Kerr-Newman singularity (the angular momentum and charge of the electron are too large for a black hole of the electron's mass, so there is no horizon) \cite{Burinskii:2005mm} (see also \cite{Schmekel:2018bcf, Burinskii:2021ipv}). The no-hair property of black holes are also similar to elementary particles: all electrons look the same. Of course our mirror model is too simple to seek certain further connections between particle physics and black holes, in particular it does not involve any charge or angular momentum. Nevertheless, \emph{precisely} because of this, it is surprising that its total emitted energy should be given by the integral of the Larmor formula.

Near-term possible theoretical applications of electron-mirror correspondence include extension to non-rectilinear trajectories; notably the uniform accelerated worldlines of Letaw \cite{Letaw:1980yv} which have Unruh-like temperatures \cite{Good:2020hav} and power distributions \cite{Good:2019aqd}. Applying the general study of Kothawala and Padmanabhan \cite{Kothawala:2009aj} to electrons moving along time-dependent accelerations and comparing the effect to an Unruh-DeWitt detector could prove fruitful for understanding thermal response. Moreover, moving mirror models can be useful in cosmology \cite{Good:2020byh}, in particular in modeling particle production due to the expansion of space \cite{castagnino}. This expansion is accelerated due to an unknown dark energy, which may not be a cosmological constant and thus can decay  \cite{Akhmedov:2009be,Polyakov:2009nq}. If dark energy is some kind of vacuum energy, it might subject to further study from mirror analogs just like Casimir energy. (In fact, dark energy could be Casimir-like \cite{ulf}.)

Near-term possible experimental applications of electron-mirror holography include leveraging the correspondence to disentangle effects in experiments like the Analog Black Hole Evaporation via Lasers (AnaBHEL) collaboration \cite{AnaBHEL:2022sri} and the RDK II collaboration \cite{nico,RDKII:2016lpd}, see also \cite{Lynch:2022rqx}.  The former exploits the accelerating relativistic moving mirror as a probe of the spectrum of quantum vacuum radiation \cite{Chen:2020sir,Chen:2015bcg} and the later measures the photon spectrum with high precision as the electron-mirror is subjected to extreme accelerations during the process of radiative neutron beta decay.
\begin{acknowledgments}
MG thanks the FY2021-SGP-1-STMM Faculty Development Competitive Research Grant No. 021220FD3951 at Nazarbayev University.        
YCO thanks the National Natural Science Foundation of China (No. 11922508) for funding support. 
\end{acknowledgments}

\bibliography{main} 
\end{document}